\documentclass[10pt]{article}
\usepackage{amsfonts}
\newcommand{\eq}{\begin{equation}}
\newcommand{\eqn}[1]{\label{#1}\end{equation}}
\newcommand{\eea}{\end{eqnarray}}
\newcommand{\eqa}{\begin{eqnarray}}
\newcommand{\eqan}[1]{\label{#1}\end{eqnarray}}
\newcommand{\ba}{\begin{array}}
\newcommand{\ea}{\end{array}}
\newcommand{\eqac}{\begin{equation}\begin{array}{rcl}}
\newcommand{\eqacn}[1]{\end{array}\label{#1}\end{equation}}

\newcommand{\text}{\rm}

\begin{document}

\title{\textbf{Gribov as a Phase Transition}}

\author{L. C. Q. Vilar$^{a}$, O. S. Ventura$^{b}$, V. E. R. Lemes$^{a}$
\footnote{lcqvilar@gmail.com, ozemar@ifes.edu.br, vitor@dft.if.uerj.br}\\
{ \small \em $^a$Instituto de F\'\i sica, Universidade do Estado do Rio de Janeiro},  \\
\small\em $ $Rua S\~{a}o Francisco Xavier 524, Maracan\~{a}, Rio de Janeiro - RJ, 20550-013, Brazil \\
\small\em $^b$Coordenadoria de F\'{\i}sica, Instituto Federal do Esp\'{\i}rito Santo,\\
\small\em Avenida Vit\'{o}ria 1729 - Jucutuquara, Vit\'{o}ria - ES, 29040 - 333, Brazil \\}
\bigskip
\maketitle

\vspace{-1cm}
\begin{abstract}

Our goal will be the description of a theory of Gribov's type as a physical process of phase
transition in the context of a spontaneous symmetry breaking. We mainly focus at the quantum stability of
the whole process.

\end{abstract}
\setcounter{page}{0}\thispagestyle{empty}

\vfill\newpage\ \makeatother

\section{Introduction}

Gribov's  ambiguity for gauge theories \cite{gribov} has always been cited in textbooks of
quantum field theory as an intriguing but inconclusive question \cite{weinberg}.
It was only after the work of Zwanziger \cite{zwanziger1,zwanziger2} that this academic problem turned into a viable path
to confinement of gauge particles \cite{all}.
The main point lies in the fact that implementing
Gribov's condition in a Yang-Mills lagrangian  takes gluons out of the physical spectrum, as a result of the breaking of
positivity of the gluon propagator \cite{all,notp}. And finally, the nonlocal
difficulty imposed by this construction was overcome after the work of Sorella et all \cite{sobreiro,dudal,capri,nele,dudal2}.
 In this environment, we will use the mathematical need of imerging Zwanziger's original theory in a wider one through
its coupling to external sources as explained in \cite{dudal}.
This mechanism is well known in the BRST approach, particularly for a theory with a soft breaking in a fundamental symmetry
for the renormalization process. The standard procedure is to introduce the
breaking itself in the starting action by coupling it to external sources in BRST doublets \cite{livro,thibes}.
In Zwanziger's case, the broken symmetry in the localized action is BRST itself.
In order to assure renormalizability, we have to study the space of counterterms
of the wider theory, including all trivial terms involving the sources, because the original theory is only recovered when
we take the sources as constants again,
what was called as a requirement  to ``attain their physical values''
\cite{zwanziger2,zwanziger3,eu}.

Then, at this moment when the sources attain their physical values, the presence of new terms required by the stability
condition derived from the BRST quantization can bring severe deformations to the theory.
For instance, propagators (including their poles) of the original theory can be disfigured, and the initial aim of
describing a particular effect lost \cite{eu,tedesco}.
It is interesting to observe that the final theory, after the inclusion of all terms required by the BRST stability,
can even be stable in the BRST sense, but yet its physics be destroyed. Obviously, explicit Feynman graphs
calculations performed starting from the classical action show the same instability, demanding the introduction of the same terms in
the original action in order to reabsorb divergences \cite{tedesco}.

What we intend to do here is to explore the mathematical procedure of immersion of Zwanziger's theory from a new angle.
We first need to compare the theory before and after we take the physical values for the sources. The starting theory, defined in the euclidean,
still in the wider space in the presence of the sources is given by \cite{schaden,ver1}
\begin{eqnarray}
S &=& \int d^{4}x_{E} \lbrace \frac{1}{4}F^{a\mu\nu}F^{a}_{\mu\nu} + i b^{a}\partial^{\mu}A^{a}_{\mu}
+ \overline{c}^{a}\partial^{\mu}D^{ab}_{\mu}c^{b}
+ \overline{\phi}^{ac}_{\nu}\partial^{\mu}D^{ab}_{\mu}\phi^{bc}_{\nu}
- \overline{\eta}^{ac}_{\nu}\partial^{\mu}D^{ab}_{\mu}\eta^{bc}_{\nu} \nonumber \\
&-& g\partial^{\mu}\overline{\eta}^{ac}_{\nu}f^{abm}(D^{be}_{\mu}c^{e})\phi^{mc}_{\nu}
- \overline{J}^{ac}_{\mu\nu}D^{ab}_{\mu}\phi^{bc}_{\nu} + J^{ac}_{\mu\nu}D^{ab}_{\mu}\overline{\phi}^{bc}_{\nu}
+ \overline{Q}^{ac}_{\mu\nu}D^{ab}_{\mu}\eta^{bc}_{\nu} - Q^{ac}_{\mu\nu}D^{ab}_{\mu}\overline{\eta}^{bc}_{\nu} \nonumber \\
&-& \overline{Q}^{ac}_{\mu\nu}gf^{adb}(D^{de}_{\mu}c^{e})\phi^{bc}_{\nu} -
J^{ac}_{\mu\nu}gf^{adb}(D^{de}_{\mu}c^{e})\overline{\eta}^{bc}_{\nu} - \overline{J}^{ac}_{\mu\nu}J^{ac}_{\mu\nu}
+ \overline{Q}^{ac}_{\mu\nu}Q^{ac}_{\mu\nu} \rbrace .
\label{gribov}
\end{eqnarray}
As we mentioned before, it is assured by the BRST procedure itself that all fields and sources introduced
beyond those needed by the pure Yang-Mills action are in doublets \cite{livro}. Let us explain this point in detail.
The BRST transformations for this theory read
\begin{eqnarray}
s A_{\mu}^{a} &=& -D_{\mu}^{ab}c^{b} = - (\partial_{\mu}\delta^{ab} + gf^{acb}A_{\mu}^{c})c^{b},
\hspace{.5cm}s c^{a}= \frac{g}{2}f^{abc}c^{b}c^{c};\nonumber  \\
s \overline{c}^{a} &=& i b^{a}, \hspace{.5cm}s b^{a}=0;\nonumber  \\
s {\phi}_{\mu}^{ab} &=& {\eta}_{\mu}^{ab} ,
\hspace{.5cm}s {\eta}_{\mu}^{ab}= 0 ; \nonumber  \\
s \overline \eta_{\mu}^{ab} &=& \overline \phi_{\mu}^{ab} ,
\hspace{.5cm}s \overline \phi_{\mu}^{ab} = 0 ;\nonumber \\
s {J}_{\mu\nu}^{ab} &=& {Q}_{\mu\nu}^{ab} ,
\hspace{.5cm}  s {Q}_{\mu\nu}^{ab} = 0 ,\nonumber  \\
s \overline{Q}_{\mu\nu}^{ab} &=& \overline{J}_{\mu\nu}^{ab} ,
\hspace{.5cm}  s \overline{J}_{\mu\nu}^{ab} = 0.
\label{dois}
\end{eqnarray}
Now, there is the well known theorem of BRST cohomology which states that all fields which are in BRST
doublets do not contribute to the physical observables of the theory \cite{livro}.
By BRST doublet it is meant the pair of fields ($\rho,\sigma$) which transform in the following structure
\begin{eqnarray}
s_{l} \rho = \sigma,  \  \  \  \  \  \    s_{l} \sigma = 0;
\label{BRST0}
\end{eqnarray}
where $s_{l}$ is just the linear part in the quantum fields of the full BRST transformation. A straightforward 
analysis of (\ref{dois}) shows that
the set (${\phi},{\eta}$), ($ \overline{\eta} , \overline{\phi}$), ($J,Q$), ($\overline Q,\overline J$) 
is formed by doublet fields, leading to the conclusion
that the fields $\overline{\phi},\phi, \eta, \overline{\eta}$ and the sources 
$J,  \overline J, Q, \overline Q$ do not integrate the Hilbert space of the theory. What is left in
eqs. (\ref{gribov}) and ({\ref{dois}) is the usual structure of a Yang-Mills theory.

Then, despite its complicated appearance, this action still has the physical content completely equivalent to that of
pure Yang-Mills \cite{versi}.
This means that up to this stage of the theory the gluon propagation obtained from (\ref{gribov})
is still that usual from Yang-Mills.

Only at the end of the BRST quantization, when we make the sources
$\overline{J}^{ac}_{\mu\nu}$, $J^{ac}_{\mu\nu}$, $\overline{Q}^{ac}_{\mu\nu}$, $Q^{ac}_{\mu\nu}$ in (\ref{gribov})
take the constant values of the original theory, i.e. $\overline{J}^{ac}_{\mu\nu}=i\gamma\delta^{ac}\delta_{\mu\nu}$,
$J^{ac}_{\mu\nu}=-i\gamma\delta^{ac}\delta_{\mu\nu}$, $\overline{Q}^{ac}_{\mu\nu}=0$, $Q^{ac}_{\mu\nu}=0$,
that we recover a different propagator for the gluon, of the Gribov's type,
\begin{equation}
<A_{\mu}^{a}A_{\nu}^{b}> = \delta^{ab}\frac{k^{2}}{k^{4}+\gamma^{4}}(\delta_{\mu\nu}-\frac{k_{\mu}k_{\nu}}{k^{2}}),
\label{gribov-original}
\end{equation}
where $\gamma$ is fixed by the Zwanziger gap equation \cite{zwanziger1,zwanziger2},
\begin{equation}
 \frac{\delta\Gamma}{\delta\gamma}=0.
\end{equation}

We note that the positivity of the gluon propagator is then lost, implying the absence of asymptotic states.
This is the moment when we go away from a conventional description of Yang-Mills for QCD,
and we come near to the Gribov-Zwanziger  point of view of a confining theory.
We realize now how the fixation process of the sources is responsible for a modification of the physics described
by the action under analysis. However, following the canonical understanding, this procedure is merely a mathematical
operation. What we envisage to do from this point on is to give a dynamical physical context to it,
as a phase transition process.

This paper is organized as follows. In section 2, we construct the symmetry breaking lagrangian leading to a Gribov propagator.
All relevant symmetries are displayed, and finally we show the most general BRST invariant cocycle
that can contribute to the gluon propagation after the phase transition. Section 3 is devoted to the analysis
of the phase transition itself. We explicitly calculate the propagators for a broken $SU(2)$ gauge group, well
suited for a comparison with lattice results. In the end we show an interesting fit with recent developments in the lattice.
In the conclusion, we summarize our work.

\section{The Symmetry Breaking Action}

We start this section reinforcing the crucial role played by the external sources in the action (\ref{gribov}). When they are
mathematically tuned into constant values, a conventional gluon propagator is conveniently converted into the Gribov propagator (\ref{gribov-original}).
We want now to replace this mathematical procedure by a natural physical process. In fact, there is a physical situation where a field
can naturally be driven to a constant value. This happens in a phase transition as a result of a symmetry breaking process
which, in general, is described by scalar fields in Landau-Ginsburg lagrangians. Then, our first conclusion is that
the mathematical role played by the sources $J$ and $\overline J$ in (\ref{gribov}) should now be physically played by complex scalar fields
${\varphi}$ and $\overline{\varphi}$. As we intend to remain as close as possible to the Zwanziger-Sorella scenario of Gribov's theory, we will also introduce anti-comuting fields
$\psi$  and $\overline{\psi}$ to play the role of $Q$ and $\overline Q$ transforming in BRST doublets with $\phi$ and $\overline {\phi}$. They will form the quartet structure

\begin{eqnarray}
s \varphi^{a} &=& \psi^{a} + g f^{abc}c^{b}\varphi^{c},
\hspace{1cm}s \psi^{a} = g f^{abc}c^{b}\psi^{c}; \nonumber  \\
s\overline{\psi}^{a} &=& \overline{\varphi}^{a} + g f^{abc}c^{b}\overline{\psi}^{c},
\hspace{1cm}s\overline{\varphi}^{a}= g f^{abc}c^{b}\overline{\varphi}^{c}.
\label{BRST2}
\end{eqnarray}

In this way, following the BRST theorem for doublet fields, we preserve the property that the theory before the phase transition
is purely Yang-Mills, without the presence of any extra degrees of freedom. Notice also that all these fields now transform in the adjoint of the  gauge
group (originally, $J$ an $Q$ in (\ref{gribov}) do no take values in the gauge group \cite{schaden},\cite{ver1}),
as it is required by their coupling to the gauge field.

The task of building a symmetry breaking theory with such a field content becomes easier after the work of K. Fujikawa \cite{fuji}.
There, such a theory was constructed for the first time as an example to study the spontaneous breaking of BRST.
Here it will serve us for the same objective, the main difference being that the fields in (\ref{BRST2}) are Lie algebra valued. Its action is given by
\begin{equation}
S_{F}= \int d^{4}x \lbrace \partial_{\mu}\overline{\varphi}\partial_{\mu}\varphi - \partial_{\mu}\overline{\psi}\partial_{\mu}\psi
- m^{2}(\overline{\varphi}\varphi - \overline{\psi}\psi ) + \frac{\lambda}{2}(\overline{\varphi}\varphi - \overline{\psi}\psi )^{2} \rbrace .
\label{fujiS}
\end{equation}
This is the symmetry breaking sector that we need to couple to the action (\ref{gribov}). As we will show in the next section, action (\ref{fujiS})
will allow the development of a non-vanishing vacuum expectation value for ${\varphi}\overline{\varphi}$, which will be the promised physical process replacing the mathematical
one in the Zwanziger-Sorella scheme.

Finally, before showing the complete new action with these improvements, there is still one last point that we want to call attention to. The tensorial nature of the sources $J$ and $Q$ in (\ref{gribov}) is lost in the change for the scalar fields of (\ref{BRST2}). In this matter, our guidance is the necessary structure needed after the symmetry breaking in order to generate a gluon propagation of Gribov's type. Also, the naive substitution of this sources  by the quartet of fields of (\ref{BRST2}) would certainly lead to non-invariant actions under BRST. We inevitably need to adapt some of the
couplings in action (\ref{gribov}) to incorporate the gauge covariance of the fields in (\ref{BRST2}). Taking all this into account, and seeking a minimal change in (\ref{gribov}), we propose
a starting action

\begin{eqnarray}
\Sigma &=& \int d^{4}x_{E}\lbrace \frac{1}{4}F_{\mu\nu}^{a}F_{\mu\nu}^{a} + i b^{a}\partial_{\mu}A_{\mu}^{a}
+ \overline{c}^{a}\partial_{\mu}D_{\mu}^{ab}c^{b} + D_{\nu}^{ab}\overline{e}_{\mu}^{b}D_{\nu}^{ac}e_{\mu}^{c}
- D_{\nu}^{ab}\overline{\omega}_{\mu}^{b}D_{\nu}^{ac}\omega_{\mu}^{c} \nonumber \\
&+& a_{2}(\overline{\varphi}^{a}\varphi^{a}-\overline{\psi}^{a}\psi^{a})(A_{\mu}^{b}(\overline{e}_{\mu}^{b}-e_{\mu}^{b}))
+ a_{2}\overline{\psi}^{a}\varphi^{a}(\partial_{\mu}c^{b})(\overline{e}_{\mu}^{b}- e_{\mu}^{b})\nonumber \\
&+& a_{2}\overline{\psi}^{a}\varphi^{a}A_{\mu}^{b}\omega_{\mu}^{b}
+ a_{3}(\overline{\varphi}^{a}\varphi^{a}-\overline{\psi}^{a}\psi^{a})
(\overline{e}_{\mu}^{b}e_{\mu}^{b} - \frac{1}{2}\overline{e}_{\mu}^{b}\overline{e}_{\mu}^{b} - \frac{1}{2}e_{\mu}^{b}e_{\mu}^{b})
\nonumber \\
&-& a_{3}\overline{\psi}^{a}\varphi^{a}\omega_{\mu}^{b}(\overline{e}_{\mu}^{b}- e_{\mu}^{b})
+ a_{4}(\overline{\varphi}^{a}\varphi^{a}-\overline{\psi}^{a}\psi^{a})A_{\mu}^{b}A_{\mu}^{b}
+ 2 a_{4}\overline{\psi}^{a}\varphi^{a}(\partial_{\mu}c^{b})A_{\mu}^{b} \nonumber \\
&+& a_{5}\mu^{2}(\overline{e}_{\mu}^{a}e_{\mu}^{a}-\overline{\omega}_{\mu}^{a}\omega_{\mu}^{a})
+ a_{6}(\overline{e}_{\mu}^{a}e_{\mu}^{a}-\overline{\omega}_{\mu}^{a}\omega_{\mu}^{a})
(\overline{e}_{\mu}^{a}e_{\mu}^{a}-\overline{\omega}_{\mu}^{a}\omega_{\mu}^{a})
+ a_{7}(\overline{\varphi}^{a}\varphi^{a}-\overline{\psi}^{a}\psi^{a})(\overline{e}_{\mu}^{b}e_{\mu}^{b}-\overline{\omega}_{\mu}^{b}\omega_{\mu}^{b})\nonumber \\
&+& D_{\mu}^{ab}\overline{\varphi}^{b}D_{\mu}^{ac}\varphi^{c} - D_{\mu}^{ab}\overline{\psi}^{b}D_{\mu}^{ac}\psi^{c}
+ \mu^{2}(\overline{\varphi}^{a}\varphi^{a}-\overline{\psi}^{a}\psi^{a})
+ \frac{\lambda}{2}(\overline{\varphi}^{a}\varphi^{a}-\overline{\psi}^{a}\psi^{a})^{2} \nonumber \\
&-& \Omega_{\mu}^{a}D_{\mu}^{ab}c^{b} + \frac{g}{2}f^{abc}L^{a}c^{b}c^{c}
+ \overline{\omega}_{\mu}^{\ast a}(\overline{e}_{\mu}^{a} + g f^{abc}c^{b}\overline{\omega}_{\mu}^{c})
+\overline{e}_{\mu}^{\ast a}(g f^{abc}c^{b}\overline{e}_{\mu}^{c}) \nonumber \\
&+& e_{\mu}^{\ast a}(\omega_{\mu}^{a} + g f^{abc}c^{b}e_{\mu}^{c})
+ \omega_{\mu}^{\ast a}(g f^{abc}c^{b}\omega_{\mu}^{c})
+ \overline{\psi}^{\ast a}(\overline{\varphi}^{a} + g f^{abc}c^{b}\overline{\psi}^{c})
+ \overline{\varphi}^{\ast a}( g f^{abc}c^{b}\overline{\varphi}^{c}) \nonumber \\
&+& \varphi^{\ast a}(\overline{\psi}^{a} + g f^{abc}c^{b}\varphi^{c})
+ \psi^{\ast a}(g f^{abc}c^{b}\psi^{c}) \rbrace .
\label{action}
\end{eqnarray}
We list the dimensions and ghost numbers of all fields of (\ref{action}):

\begin{table}[h]
\begin{center}
\begin{tabular}{|l|c|c|c|c|c|c|c|c|c|c|c|c|c|c|c|c|c|c|c|c|c|c|}
\hline
& $A$ & $b$ & $c$ & $\overline{c}$ & $e $ & $\overline{e}
$ & $\omega $ & $\overline{\omega}$ & $\varphi $ & $\overline{\varphi}$ & $\psi $ & $\overline{\psi}$
\\ \hline
dimension & 1 & 2 & 0 & 2 & 1 & 1 & 1 & 1 & 1 & 1 & 1 & 1
\\ \hline
ghost number
& 0 & 0 & 1 & $-1$ & 0 & 0 & 1 & $-1$ & 0 & 0 & 1 & $-1$  \\ \hline
\end{tabular}
\caption{The quantum numbers of fields and sources of the theory}
\label{table}
\end{center}
\end{table}

Now, let us spend a few words on the nature of this action (\ref{action}), specially beyond the usual terms of Yang-Mills and Faddev-Popov gauge fixing. The main point to be stressed once more is that, analogously to action (\ref{gribov}), the extra terms do not comprise any new physics. To prove this, we just need to take a look at the full set of BRST transformations which leave action (\ref{action}) invariant, which is compounded by transformations (\ref{BRST2}) together with

\begin{eqnarray}
s A_{\mu}^{a} &=& -D_{\mu}^{ab}c^{b} = - (\partial_{\mu}\delta^{ab} + gf^{acb}A_{\mu}^{c})c^{b},
\hspace{.5cm}s c^{a}= \frac{g}{2}f^{abc}c^{b}c^{c};\nonumber  \\
s \overline{c}^{a} &=& i b^{a}, \hspace{.5cm}s b^{a}=0;\nonumber  \\
s e_{\mu}^{a} &=& \omega_{\mu}^{a} + g f^{abc}c^{b}e_{\mu}^{c},
\hspace{.5cm}s \omega_{\mu}^{a} = g f^{abc}c^{b}\omega_{\mu}^{c}; \nonumber  \\
s\overline{\omega}_{\mu}^{a} &=& \overline{e}_{\mu}^{a} + g f^{abc}c^{b}\overline{\omega}_{\mu}^{c},
\hspace{.5cm}s\overline{e}_{\mu}^{a}= g f^{abc}c^{b}\overline{e}_{\mu}^{c}.
\label{BRST}
\end{eqnarray}

The doublets ($ \varphi,  \psi$), ($\overline{\psi}, \overline{\varphi}$), ($e,\omega$) and ($\overline{\omega},\overline{e}$) are easily identified, and leave only the traditional Yang-Mills observables in the physical spectrum.

The BRST sources  $ \Omega$, $L$, $\overline{\omega}^{\ast}$, $\overline{e}^{\ast}$, ${e}^{\ast }$, ${\omega}^{\ast}$, $\overline{\psi}^{\ast}$, $\overline{\varphi}^{\ast}$, ${\varphi}^{\ast}$ and ${\psi}^{\ast}$
appear in (\ref{action}) coupled to the non-linear sectors of the BRST transformations of their respective fields
in (\ref{BRST}). This is standard in the BRST renormalization procedure and accounts for the renormalization
of the transformations themselves \cite{livro}.

We have also anticipated in (\ref{action}) the presence of the elements with coefficients $a_3$, $a_ 4$, $a_5$, $a_6$ and $a_7$,
which are demanded after the BRST renormalization in order to make the theory quantically stable. Terms of this nature
also appear for the same reason after the renormalization of action (\ref{gribov}), and lead to important improvements
on the Gribov propagator \cite{dudal-nele}.

We proceed now with the BRST renormalization by observing that the BRST operator defined in (\ref{BRST2}) and (\ref{BRST}) is nilpotent, i.e.,

\begin{eqnarray}
s^2\theta =0,
\label{BRST1}
\end{eqnarray}
where $\theta$ stands for all the fields and sources of the theory. In functional form, this implies the Slavnov-Taylor identity

\begin{eqnarray}
S(\Sigma ) &=& \int d^{4}x_{E}\lbrace \frac{\delta\Sigma}{\delta A_{\mu}^{a}}\frac{\delta\Sigma}{\delta\Omega_{\mu}^{a}}
+ \frac{\delta\Sigma}{\delta c^{a}}\frac{\delta\Sigma}{\delta L^{a}}
+ \frac{\delta\Sigma}{\delta \overline{\omega}_{\mu}^{a}}\frac{\delta\Sigma}{\delta \overline{\omega}_{\mu}^{\ast a}}
+ \frac{\delta\Sigma}{\delta \overline{e}_{\mu}^{a}}\frac{\delta\Sigma}{\delta \overline{e}_{\mu}^{\ast a}}
+ \frac{\delta\Sigma}{\delta e_{\mu}^{a}}\frac{\delta\Sigma}{\delta e_{\mu}^{\ast a}}
+ \frac{\delta\Sigma}{\delta \omega_{\mu}^{a}}\frac{\delta\Sigma}{\delta \omega_{\mu}^{\ast a}} \nonumber \\
&+&\frac{\delta\Sigma}{\delta \overline{\psi}^{a}}\frac{\delta\Sigma}{\delta \overline{\psi}^{\ast a}}
+ \frac{\delta\Sigma}{\delta \overline{\varphi}^{a}}\frac{\delta\Sigma}{\delta \overline{\varphi}^{\ast a}}
+ \frac{\delta\Sigma}{\delta \varphi^{a}}\frac{\delta\Sigma}{\delta \varphi^{\ast a}}
+ \frac{\delta\Sigma}{\delta \psi^{a}}\frac{\delta\Sigma}{\delta \psi^{\ast a}}
+ i b^{a}\frac{\delta\Sigma}{\delta\overline{c}^{a}} \rbrace =0,
\label{Slavnov}
\end{eqnarray}
and the nilpotency of the linearized Slavnov-Taylor operator

\begin{eqnarray}
B^2_{\Sigma}= 0, \nonumber \\
\label{linearizado}
\end{eqnarray}
\begin{eqnarray}
B_{\Sigma} &=&\int d^{4}x_{E}\lbrace \frac{\delta\Sigma}{\delta A_{\mu}^{a}}\frac{\delta}{\delta\Omega_{\mu}^{a}}
+ \frac{\delta\Sigma}{\delta c^{a}}\frac{\delta}{\delta L^{a}}
+ \frac{\delta\Sigma}{\delta \overline{\omega}_{\mu}^{a}}\frac{\delta}{\delta \overline{\omega}_{\mu}^{\ast a}}
+ \frac{\delta\Sigma}{\delta \overline{e}_{\mu}^{a}}\frac{\delta}{\delta \overline{e}_{\mu}^{\ast a}}
+ \frac{\delta\Sigma}{\delta e_{\mu}^{a}}\frac{\delta}{\delta e_{\mu}^{\ast a}}
+ \frac{\delta\Sigma}{\delta \omega_{\mu}^{a}}\frac{\delta}{\delta \omega_{\mu}^{\ast a}} \nonumber \\
&+&\frac{\delta\Sigma}{\delta\Omega_{\mu}^{a}}\frac{\delta}{\delta A_{\mu}^{a}}
+ \frac{\delta\Sigma}{\delta L^{a}} \frac{\delta}{\delta c^{a}}
+ \frac{\delta\Sigma}{\delta \overline{\omega}_{\mu}^{\ast a}}\frac{\delta}{\delta \overline{\omega}_{\mu}^{a}}
+ \frac{\delta\Sigma}{\delta \overline{e}_{\mu}^{\ast a}}\frac{\delta}{\delta \overline{e}_{\mu}^{a}}
+ \frac{\delta\Sigma}{\delta e_{\mu}^{\ast a}} \frac{\delta}{\delta e_{\mu}^{a}}
+ \frac{\delta\Sigma}{\delta \omega_{\mu}^{\ast a}}\frac{\delta}{\delta \omega_{\mu}^{a}}\nonumber \\
&+&\frac{\delta\Sigma}{\delta \overline{\psi}^{a}}\frac{\delta}{\delta \overline{\psi}^{\ast a}}
+ \frac{\delta\Sigma}{\delta \overline{\varphi}^{a}}\frac{\delta}{\delta \overline{\varphi}^{\ast a}}
+ \frac{\delta\Sigma}{\delta \varphi^{a}}\frac{\delta}{\delta \varphi^{\ast a}}
+ \frac{\delta\Sigma}{\delta \psi^{a}}\frac{\delta}{\delta \psi^{\ast a}}
+ i b^{a}\frac{\delta}{\delta\overline{c}^{a}}\nonumber \\
&+&\frac{\delta\Sigma}{\delta \overline{\psi}^{\ast a}}\frac{\delta}{\delta \overline{\psi}^{a}}
+ \frac{\delta\Sigma}{\delta \overline{\varphi}^{\ast a}}\frac{\delta}{\delta \overline{\varphi}^{ a}}
+ \frac{\delta\Sigma}{\delta \varphi^{\ast a}}\frac{\delta}{\delta \varphi^{a}}
+ \frac{\delta\Sigma}{\delta \psi^{\ast a}}\frac{\delta}{\delta \psi^{a}}
 \rbrace .\nonumber
\end{eqnarray}

Following the general prescription of BRST, we need to find all counterterms $\Sigma_{c}$ with UV dimension up to 4, zero ghost number and invariant under the action of $B_{\Sigma}$, i.e.,

\begin{equation}
\Gamma = \Sigma + \hbar\Sigma_{c} ,\hspace{.5cm}
B_{\Sigma}\Sigma_{c} = 0 . \nonumber
\label{Slavnov1}
\end{equation}

Usually, this condition (\ref{Slavnov1}) alone is not sufficiently restrictive on the most general form for the counterterm. We must find more conditions in order to reduce the number of independent elements in $\Sigma_{c}$. These conditions are obtained from classical symmetries, possibly linearly broken in the quantum fields, satisfied by the classical action and compatible with the Quantum Action Principle \cite{livro}. This last demand will ensure that such symmetries will be associated to quantum Ward identities in fact restricting $\Sigma_{c}$ \cite{lowenstein}. Among them, the most useful identities are

\begin{enumerate}
\item The gauge fixing and antighost equation
\begin{eqnarray}
\frac{\delta\Sigma}{\delta b^{a}} &=& i\partial_{\mu}A_{\mu}^{a}, \nonumber \\
\frac{\delta\Sigma}{\delta\overline{c}^{a}}+\partial_{\mu}\frac{\delta\Sigma}{\delta\Omega_{\mu}^{a}}&=&0.
\label{calibre}
\end{eqnarray}
\item The ghost equation
\begin{eqnarray}
&G^{a}(\Sigma)&= \int d^{4}x_{E}\lbrace \frac{\delta\Sigma}{\delta c^{a}} -ig f^{abc}\overline{c}^{b}\frac{\delta\Sigma}{\delta b^{c}}\rbrace \\
&=&\hspace{-.7cm}- gf^{abc}\hspace{-.1cm}\int \hspace{-.1cm}d^{4}x_{E}\lbrace A_{\mu}^{b}\Omega_{\mu}^{c}\hspace{-.05cm}
+\hspace{-.05cm}L^{b}c^{c}
+ \overline{\omega}_{\mu}^{\ast b}\overline{\omega}_{\mu}^{ c}
+ \omega_{\mu}^{\ast b}\omega_{\mu}^{ c}
- \overline{e}_{\mu}^{\ast b}\overline{e}_{\mu}^{ c}
- e_{\mu}^{\ast b}e_{\mu}^{ c}
+ \overline{\psi}^{\ast b}\overline{\psi}^{c}
+ \psi^{\ast b}\psi^{ c}
- \overline{\varphi}^{\ast b}\overline{\varphi}^{ c}
- \varphi^{\ast b}\varphi^{ c} \rbrace. \nonumber
\label{eq-ghost}
\end{eqnarray}
\item The localization fields equation
\begin{eqnarray}
L_{\mu\nu}(\Sigma)&=&\int d^{4}x_{E}\lbrace \overline{e}_{\nu}^{a}\frac{\delta\Sigma}{\delta\overline{\omega}_{\mu}^{a}}
+ \overline{\omega}_{\mu}^{\ast a}\frac{\delta\Sigma}{\delta\overline{e}_{\nu}^{\ast a}}
+ \omega_{\mu}^{a}\frac{\delta\Sigma}{\delta e_{\nu}^{a}}
- e_{\nu}^{\ast a}\frac{\delta\Sigma}{\delta \omega_{\mu}^{\ast a}}
+ \omega_{\mu}^{a}\frac{\delta\Sigma}{\delta \overline{e}_{\nu}^{a}}
- \overline{e}_{\nu}^{\ast a}\frac{\delta\Sigma}{\delta \omega_{\mu}^{\ast a}}
+ e_{\nu}^{a}\frac{\delta\Sigma}{\delta\overline{\omega}_{\mu}^{a}}
+ \overline{\omega}_{\mu}^{\ast a}\frac{\delta\Sigma}{\delta e_{\nu}^{\ast a}} \rbrace  \nonumber \\
&=&\int d^{4}x_{E}\lbrace \overline{\omega}_{\mu}^{\ast a}\omega_{\nu}^{a} \rbrace.
\label{eq-loc}
\end{eqnarray}
\item The matter fields equation
\begin{eqnarray}
M(\Sigma)&=&\int d^{4}x_{E}\lbrace \overline{\varphi}^{a}\frac{\delta\Sigma}{\delta\overline{\varphi}^{a}}
- \overline{\varphi}^{\ast a}\frac{\delta\Sigma}{\delta\overline{\varphi}^{\ast a}}
- \varphi^{a}\frac{\delta\Sigma}{\delta\varphi^{a}}
+ \varphi^{\ast a}\frac{\delta\Sigma}{\delta\varphi^{\ast a}}
+ \overline{\psi}^{a}\frac{\delta\Sigma}{\delta\overline{\psi}^{a}}
- \overline{\psi}^{\ast a}\frac{\delta\Sigma}{\delta\overline{\psi}^{\ast a}}
- \psi^{a}\frac{\delta\Sigma}{\delta\psi^{a}}
+ \psi^{\ast a}\frac{\delta\Sigma}{\delta\psi^{\ast a}} \rbrace \nonumber \\
&=& \int d^{4}x_{E}\lbrace \overline{\psi}^{\ast a}\overline{\varphi}^{a} - \overline{\varphi}^{\ast a}\overline{\psi}^{a}
\rbrace.
\end{eqnarray}
\item The localization antighost equation
\begin{equation}
I(\Sigma) = \int d^{4}x_{E}\lbrace \omega_{\mu}^{a}\frac{\delta\Sigma}{\delta \overline{\omega}_{\nu}^{a}} +
\omega_{\nu}^{a}\frac{\delta\Sigma}{\delta \overline{\omega}_{\mu}^{a}} -
\overline{\omega}_{\mu}^{\ast a}\frac{\delta\Sigma}{\delta \omega_{\nu}^{\ast a}} -
\overline{\omega}_{\nu}^{\ast a}\frac{\delta\Sigma}{\delta \omega_{\mu}^{\ast a}} \rbrace =0.
\end{equation}
\item The localization ghost equation
\begin{eqnarray}
T(\Sigma) &=& \int d^{4}x_{E}\lbrace \overline{\psi}^{a}\frac{\delta\Sigma}{\delta\psi^{a}}
- \psi^{\ast a}\frac{\delta\Sigma}{\delta\overline{\psi}^{\ast a}} \rbrace \nonumber \\
&=& \int d^{4}x_{E}\lbrace \varphi^{\ast a}\overline{\psi}^{a} - \psi^{\ast a}\overline{\varphi}^{a} \rbrace .
\label{ghostpsi}
\end{eqnarray}
\end{enumerate}
These symmetries, compatible with the QAP, lead to the following constraints on $\Sigma_{c}$:

\begin{eqnarray}
\frac{\delta\Sigma_{c}}{\delta b^{a}}&=&0,\:\:
G^{a}(\Sigma_{c}) = 0; \nonumber \\
L_{\mu\nu}(\Sigma_{c})&=&0 ,\:\:
M(\Sigma_{c})=0; \nonumber \\
I(\Sigma_{c})&=&0,\:\:
T(\Sigma_{c})=0.
\label{counterterm}
\end{eqnarray}

As we have mentioned earlier, to find the whole set of relevant counterterms,
we must take into account not only the nontrivial elements of the cohomology problem established by (\ref{linearizado}) and (\ref{Slavnov1}),
but also those which can be written as BRST variations, trivial in the BRST cohomology.
This is also needed here because after the symmetry breaking we will have to redefine the scalar fields in order to
expand the theory around the new true vacuum. In the end, mathematically,  the effect will be the same as the
BRST process of fixing the external sources $\overline{J}^{ac}_{\mu\nu}$, $J^{ac}_{\mu\nu}$, $\overline{Q}^{ac}_{\mu\nu}$, $Q^{ac}_{\mu\nu}$
in (\ref{gribov}), or bringing them to their \textit{physical values} in Zwanziger's formulation.
As our main concern is to observe the possible changes in the gluon propagator after the phase transition,
we will only list the counterterms that may cause this change. By this we mean all elements that can give rise
to bilinears after the shift of $\varphi$ and $\overline{\varphi}$ as constants in terms of their vacuum expectation values.
The most general cocycle of this kind, constrained by (\ref{counterterm}), is
\begin{eqnarray}
\Sigma_{c} &=& \int d^{4}x_{E}\lbrace \frac{\rho}{4}F_{\mu\nu}^{a}F_{\mu\nu}^{a}\rbrace + B_{\Sigma}\Delta  ,\nonumber \\
\Delta &=& \int d^{4}x_{E}\lbrace \sigma_{0}(\partial_{\mu}\overline{c}^{a}+\Omega_{\mu}^{a})A_{\mu}^{a}
+ \sigma_{1}(D_{\nu}\overline{\omega}_{\mu})^{a}(D_{\nu}e_{\mu})^{a} + \sigma_{2}\overline{\psi}^{a}\varphi^{a}A_{\mu}^{b}(\overline{e}_{\mu}-e_{\mu})^{b}\nonumber \\
&-& \frac{\sigma_{3}}{2}\overline{\psi}^{a}\varphi^{a}(\overline{e}_{\mu}-e_{\mu})^{b}(\overline{e}_{\mu}-e_{\mu})^{b}
+ \sigma_{4}\overline{\psi}^{a}\varphi^{a}A_{\mu}^{b}A_{\mu}^{b} + \sigma_{5}\mu^{2}\overline{\omega}_{\mu}^{a}e_{\mu}^{a}\nonumber \\
&+&\sigma_{6}\overline{\omega}_{\mu}^{a}e_{\mu}^{a}(\overline{e}_{\nu}^{b}e_{\nu}^{b} - \overline{\omega}_{\nu}^{b}\omega_{\nu}^{b})
+ \sigma_{7}\overline{\psi}^{a}\varphi^{a}(\overline{e}_{\nu}^{b}e_{\nu}^{b} - \overline{\omega}_{\nu}^{b}\omega_{\nu}^{b})\nonumber \\&+&\sigma_{8}(D_{\mu}\overline{\psi})^{a}(D_{\mu}\varphi )^{a} - \sigma_{9}\mu^{2}\overline{\psi}^{a}\varphi^{a}
+ \sigma_{10}\frac{\lambda}{2}\overline{\psi}^{a}\varphi^{a}(\overline{\varphi}^{b}\varphi^{b} - \overline{\psi}^{b}\psi^{b})\nonumber \\
&+&\sigma_{11}(\omega_{\mu}^{\ast a}\overline{\omega}_{\mu}^{a}- e_{\mu}^{\ast a}\overline{e}_{\mu}^{a}
+ \overline{\omega}_{\mu}^{\ast a}\omega_{\mu}^{a}- \overline{e}_{\mu}^{\ast a}e_{\mu}^{a})\nonumber \\
&+& \sigma_{12}(\psi^{\ast a}\overline{\psi}^{a}- \varphi^{\ast a}\overline{\varphi}^{a}
+ \overline{\psi}^{\ast a}\psi^{a}- \overline{\varphi}^{\ast a}\varphi^{a}) \rbrace .
\label{contra}
\end{eqnarray}
Here we see that, if we had not anticipated the presence of the cocycles with coefficients $a_{3}$, $a_{4}$, $a_{5}$, $a_{6}$, $a_{7}$ in the starting action (\ref{action}), then it would be vindicated now by $\sigma_{3}$, $\sigma_{4}$, $\sigma_{5}$, $\sigma_{6}$, $\sigma_{7}$ respectively, as a result of the quantum stability of the action by the BRST procedure.
As a comment on this calculation, we would like to remark the usefulness of the identities (\ref{counterterm})
in deriving (\ref{contra}) with a counter-example. If it was not for the identity $L_{\mu\nu}(\Sigma_{c})=0$, coming from
the symmetry (\ref{eq-loc}), an element as

\begin{eqnarray}
(D_{\mu}\overline e_{\mu})^{a}(D_{\nu}e_{\nu})^{a}-(D_{\mu}\overline \omega_{\mu})^{a}(D_{\nu}\omega_{\nu})^{a}
\label{teste2}
\end{eqnarray}
would be allowed by the use of (\ref{Slavnov1}) only, as this element is BRST invariant (trivial, in fact).
If such a term was to be introduced in the initial action in order to reabsorb possible divergences coming from
Feynman graph calculations, this would indicate the breaking of the stability of our classical action (\ref{action})
under quantum corrections, and, worst of all, the presence of (\ref{teste2}) would bring damages to our objective
of generating a theory with a Gribov propagator after the phase transition. The essence of the identity generated
by $L_{\mu\nu}$ just states that there are no possible divergent graphs contributing to (\ref{teste2}),
as this element does not satisfy this constraint. Other harmful BRST invariant elements would also be generated in (\ref{contra}) if the constraints (\ref{counterterm}) were not blocking them.

Finally, the counterterm action (\ref{contra}) has exactly the same cocycles already present in the original action.
This shows its stability under renormalization.

\section{Fujikawa's phase transition}

The idea now is to characterize the soft breaking of the BRST symmetry as a spontaneous symmetry breaking of the
Fujikawa's type. Following \cite{fuji}, we then rewrite the potential for the bosonic sector of the action (\ref{action}) as
\begin{equation}
V(\overline{\varphi},\varphi) = \mu^{2}\overline{\varphi}^{b}\varphi^{b}
+ \frac{\lambda}{2}(\overline{\varphi}^{b}\varphi^{b})^{2}.
\label{potential}
\end{equation}
Once the coefficient $\mu^{2}$ dinamically accquires a negative value, being understood as a chemical potential, the potential (\ref{potential})
developes a non-symmetric degenerate vacua, defined by the new minimum given by
\begin{eqnarray}
\frac{\delta V}{\delta\varphi^{a}}|_{\varphi_{0}}&=&0\hspace{.3cm}\Longrightarrow\hspace{.3cm}
-|\mu^{2}|\overline{\varphi}^{a}_{0} + \lambda\overline{\varphi}^{a}_{0}(\overline{\varphi}^{b}_{0}\varphi^{b}_{0})=0, \nonumber \\
\overline{\varphi}^{b}_{0}\varphi^{b}_{0}&=&\frac{|\mu^{2}|}{\lambda}.
\end{eqnarray}
In simple terms $\varphi^{a}$ and $\overline{\varphi}^{a}$ develop non-vanishing vacuum expectation values.
In order to identify the nature of the spectrum we need to expand the potential (\ref{potential})
around the new vacuum by redefining $\overline{\varphi}^{a}$ and $\varphi^{a}$.

In the following, we take the $SU(2)$ group as an example for the gauge group. This is the case for which we can find conclusive results
in four dimensions lattice calculations \cite{Langfeld,Amemiya,Attilio,Muller,Attilio2}. Among them, we cite that, in the Landau gauge, refined
Gribov's propagators are obtained as the gauge propagator.
More recently, simulations using larger lattices showed an interesting effect of the kind of an abelian dominance,
where the Gribov's gauge propagator associated to the Cartan Subgroup (diagonal) generator has a lighter mass
than the other propagators \cite{attilio}. We will show how this kind of effect can be accomplished in our scenario.

First we shift $\overline{\varphi}^{a}$ and $\varphi^{a}$ to their vacuum expectation values defined as:
\begin{eqnarray}
\overline{\varphi}^{a} &&\Rightarrow \overline{\varphi}^{a} + \delta^{ai}v^{i}\nonumber \\
\varphi^{a}&&\Rightarrow \varphi^{a} + \delta^{ai}v^{i}, \nonumber \\
v^{i}v^{i}&=&\frac{|\mu^{2}|}{\lambda},
\label{quebra}
\end{eqnarray}
where the latin index $i$ stands for the diagonal subgroup, which in the $SU(2)$ case means $\delta^{ai}v^{i}\equiv\delta^{a3}v^{3}$.
When substituted in (\ref{potential}), we obtain
\begin{equation}
V(\overline{\varphi},\varphi)= \frac{\lambda}{2}v^{i}v^{j}(\overline{\varphi}^{i}+\varphi^{i})(\overline{\varphi}^{j}+\varphi^{j})
+\lambda v^{i}(\overline{\varphi}^{i}+\varphi^{i})(\overline{\varphi}^{a}\varphi^{a})
+\frac{\lambda}{2}(\overline{\varphi}^{a}\varphi^{a})^{2}.
\end{equation}
This result indicates that the combination $(\overline{\varphi}^{i}-\varphi^{i})$
is massless.

Then, using (\ref{quebra}), after the spontaneous symmetry breaking, the action (\ref{action}), without the source terms irrelevant for the following discussion, achieves the form:
\begin{eqnarray}
&&\int d^{4}x_{E}\lbrace \frac{1}{4}F_{\mu\nu}^{a}F_{\mu\nu}^{a} + i b^{a}\partial_{\mu}A_{\mu}^{a}
+ \overline{c}^{a}\partial_{\mu}D_{\mu}^{ab}c^{b} + D_{\nu}^{ab}\overline{e}_{\mu}^{b}D_{\nu}^{ac}e_{\mu}^{c}
- D_{\nu}^{ab}\overline{\omega}_{\mu}^{b}D_{\nu}^{ac}\omega_{\mu}^{c} \nonumber \\
&+& a_{2}(\overline{\varphi}^{a}\varphi^{a} + v^{i}[\overline{\varphi}^{i}+\varphi^{i}] +\frac{|\mu^{2}|}{\lambda}-\overline{\psi}^{a}\psi^{a})A_{\mu}^{b}(\overline{e}_{\mu}^{b}-e_{\mu}^{b})
+ a_{2}\overline{\psi}^{a}[\varphi^{a}+\delta^{ai}v^{i}](\partial_{\mu}c^{b})(\overline{e}_{\mu}^{b}- e_{\mu}^{b})\nonumber \\
&+& a_{2}\overline{\psi}^{a}[\varphi^{a}+\delta^{ai}v^{i}]A_{\mu}^{b}\omega_{\mu}^{b}
+ a_{3}(\overline{\varphi}^{a}\varphi^{a}+ v^{i}[\overline{\varphi}^{i}+\varphi^{i}] +\frac{|\mu^{2}|}{\lambda}-\overline{\psi}^{a}\psi^{a})
(\overline{e}_{\mu}^{b}e_{\mu}^{b} - \frac{1}{2}\overline{e}_{\mu}^{b}\overline{e}_{\mu}^{b} - \frac{1}{2}e_{\mu}^{b}e_{\mu}^{b})
\nonumber \\
&-& a_{3}\overline{\psi}^{a}[\varphi^{a}+\delta^{ai}v^{i}]\omega_{\mu}^{b}(\overline{e}_{\mu}^{b}- e_{\mu}^{b})
+ a_{4}(\overline{\varphi}^{a}\varphi^{a}+ v^{i}[\overline{\varphi}^{i}+\varphi^{i}] +\frac{|\mu^{2}|}{\lambda}-\overline{\psi}^{a}\psi^{a})A_{\mu}^{b}A_{\mu}^{b} \nonumber \\
&+& 2 a_{4}\overline{\psi}^{a}[\varphi^{a}+\delta^{ai}v^{i}](\partial_{\mu}c^{b})A_{\mu}^{b}
+ a_{5}\mu^{2}(\overline{e}_{\mu}^{a}e_{\mu}^{a}-\overline{\omega}_{\mu}^{a}\omega_{\mu}^{a})
+ a_{6}(\overline{e}_{\mu}^{a}e_{\mu}^{a}-\overline{\omega}_{\mu}^{a}\omega_{\mu}^{a})
(\overline{e}_{\mu}^{a}e_{\mu}^{a}-\overline{\omega}_{\mu}^{a}\omega_{\mu}^{a}) \nonumber \\
&+& a_{7}(\overline{\varphi}^{a}\varphi^{a}+ v^{i}[\overline{\varphi}^{i}+\varphi^{i}] +\frac{|\mu^{2}|}{\lambda}-\overline{\psi}^{a}\psi^{a})(\overline{e}_{\mu}^{b}e_{\mu}^{b}-\overline{\omega}_{\mu}^{b}\omega_{\mu}^{b})\nonumber \\
&+& D_{\mu}^{ab}\overline{\varphi}^{b}D_{\mu}^{ac}\varphi^{c} - D_{\mu}^{ab}\overline{\psi}^{b}D_{\mu}^{ac}\psi^{c}
+ g f^{abc}\delta^{ci}v^{i}A_{\mu}^{b}\partial_{\mu}(\overline{\varphi}^{a}+ \varphi^{a})
+ g^{2}f^{abi}f^{adj}v^{i}v^{j}A_{\mu}^{b}A_{\mu}^{d}\nonumber \\
&+&\frac{\lambda}{2}v^{i}v^{j}(\overline{\varphi}^{i}+\varphi^{i})(\overline{\varphi}^{j}+\varphi^{j})
+\lambda v^{i}(\overline{\varphi}^{i}+\varphi^{i})(\overline{\varphi}^{a}\varphi^{a}-\overline{\psi}^{a}\psi^{a})
+ \frac{\lambda}{2}(\overline{\varphi}^{a}\varphi^{a}-\overline{\psi}^{a}\psi^{a})^{2}  \rbrace .
\label{acquebra}
\end{eqnarray}

As a final result of quantum stability and symmetry breaking, we observe the appearance of new elements not present
in the starting action (\ref{action}): there is the traditional term which will contribute explicitly as a mass term for $A_{\mu}^{a}$
after the phase transition with coefficient $g^{2}$; a term which will also contribute explicitly as a mass term for $A_{\mu}^{a}$ with coefficient $a_{4}\frac{|\mu^{2}|}{\lambda}$; terms proportional to $a_{3}$, $a_{5}$ and $a_{7}$,
which can also be found in the Zwanziger modified model \cite{final}; and a new term for the localizing fields with coefficient $a_{2}$.
The repercussion of a mass term in a Gribov-Zwanziger action was firstly analysed in \cite{dudal}.
There it was shown that the Gribov's propagator is changed minimally to a Stingl propagator \cite{shifman}.
On the other side, terms mixing the localizing fields appeared in \cite{eu} and in \cite{tedesco}.
Although in different contexts, the presence of these terms, demanded by BRST stability,
ruined both constructions as a result of the deformation implemented in the original propagators.
But in the present case, one can see that the final gauge propagator obtained from (\ref{acquebra}) is still of Gribov's type.
Now, as the breaking is taken along the abelian subgroup, it is natural that we will find
different propagators for the diagonal and off diagonal gauge fields,
\begin{eqnarray}
<A_{\mu}^{3}A_{\nu}^{3}>&=& \frac{k^{2} + m^{2}}{k^{4} + M^{2}k^{2} + \gamma^{4}}(\delta_{\mu\nu}-\frac{k_{\mu}k_{\nu}}{k^{2}}),\nonumber \\
<A_{\mu}^{a}A_{\nu}^{b}>&=& (\delta^{ab}-\delta^{a3}\delta^{b3})\frac{k^{2} + m^{2}}{k^{4} + M^{2}k^{2}+ 2g^{2}\frac{|\mu^{2}|}{\lambda}k^{2}
+ \gamma^{4} + 2g^{2} m^{2} \frac{|\mu^{2}|}{\lambda}}(\delta_{\mu\nu}-\frac{k_{\mu}k_{\nu}}{k^{2}}),\nonumber \\
m^{2}&=& (2a_{3} + a_{7} + a_{5}\lambda)\frac{|\mu^{2}|}{\lambda},\hspace{.3cm}
M^{2}= 2a_{4}\frac{|\mu^{2}|}{\lambda} + m^{2} ,\hspace{.3cm}
\gamma^{4}= a_{2}^{2}\frac{|\mu^{4}|}{\lambda^{2}} + 2a_{4}m^{2}\frac{|\mu^{2}|}{\lambda}.
\label{propagador}
\end{eqnarray}
As a generalization of Gribov, the form of the previous propagators became known as the refined Gribov propagator \cite{dudal-nele}. Let us remark some relevant points.
The first one is that, from (\ref{propagador}), one can see that the coefficient $a_{2}$ is in fact responsible for the generation of the Gribov behavior (making $a_{2}$ null, turns (\ref{propagador}) into an ordinary massive propagator). The second point is that contrary to what is usual in a conventional symmetry breaking process, a mass gap $M^{2}$ appears in the diagonal direction, although it actually is not broken.
The last point is that the usual term  $g^{2}f^{abi}f^{adj}v^{i}v^{j}A_{\mu}^{b}A_{\mu}^{d}$,
coming from the symmetry breaking, is responsible for the differences between the propagators of the gauge components, which cannot be obtained in the standard Gribov-Zwanziger-Sorella scenario.
Choosing $M^{2}$ as positive from the beginning, the off diagonal propagator $<A_{\mu}^{a}A_{\nu}^{b}>$ will have
a bigger mass gap ($M^{2}+ 2g^{2}\frac{|\mu^{2}|}{\lambda}$)
than the diagonal one. This is the previously cited lattice result \cite{attilio}.

\section{Conclusion}
The idea of Gribov's confinement has been gaining more and more consistency along the last years, as its non-local
structure became tractable by Zwanziger's localization process \cite{zwanziger1} and its renormalizability
concluded within the BRST  approach \cite{zwanziger2}. Our work here is just an unpretentious step further trying
to harmonize this project with the general picture of quantum gauge theories, in particular with the Grand Unified Theories program.
In this sense, we felt it indispensable to seek for a possible description of the Gribov-Zwanziger-Sorella scheme
in a way closer to a symmetry breaking process, an essential issue of GUTs in general. Our main concern was in demanding
the renormalizability of the theory, and at the same time preserving the achievements already obtained in the GZS scheme.
This was not assured from the beginning of our work, as the introduction of propagating fields $\varphi^{a}$ and $\overline{\varphi}^{a}$ in place
of what originally were sources in GZS would possibly bring obstructions. In fact, some alternatives proved to be non-renormalizable
or else the BRST stability took the theory away from a Gribov's description. We see the final result of the quantum stable
action (\ref{contra}), its form after the symmetry breaking (\ref{acquebra}), and the refined Gribov's propagator
(\ref{propagador}) for the gluon as a sign that GZS may become a part of the GUT program. Of course, up to now we
just showed an initial compatibility of these ideas. In order to fulfill them, one should be able to develop points as
the construction of the particle spectrum in the broken phase, the agreement with successful  aspects of QCD and GUTs,
theoretical and experimental, that are already available, or even obtaining a glimpse of what could be the possible
mechanism for the quark confinement in this picture. All of this is well beyond what is shown here, but we hope that
this can become useful in the future.

\section*{Acknowledgements}
The Conselho Nacional de Desenvolvimento Cient\'{\i}ico e tecnol\'{o}gico CNPq-
Brazil, Funda\c{c}\~{a}o de Amparo a Pesquisa do Estado do Rio de Janeiro
(Faperj) and the SR2-UERJ are acknowledged for the financial support.

\end{document}